\documentstyle[11pt,newpasp,twoside,epsf]{article}
\begin{document}
\title{Molecules as Tracers of PN Structure}
\author{P. J. Huggins}
\affil{Physics Department, New York University, 4 Washington Place,
New York NY 10003}

\begin{abstract}
Molecular gas plays an important role in the structure of planetary
nebulae: it is a major component of the equatorial tori of bipolar
nebulae, it forms the cores of globules and related mircostructures,
and is the likely origin of multiple arcs. It is also a key component
during the early stages of formation where interactions with outflows
or jets provide an important shaping mechanism.

\end{abstract}

\section{Introduction}
The matter that evolves into a planetary nebula (PN) is initially
ejected during the AGB and proto-PN phases, largely in the form of
molecular gas. The structure in this gas and its response to ionizing
radiation and fast winds or jets from the central star are important
features of PN formation. In this paper we briefly describe and
comment on some recent results which focus on this structural
development.

\section{Precursor Envelopes and Multiple Shells}
We turn first to the structure of AGB envelopes which is basic for
understanding PNe since it determines the environment in which the
nebulae form. It is also crucial for understanding the mass-loss
process and has been widely discussed (e.g., Olofsson 1999). A recent,
valuable contribution on this issue is the atlas of circumstellar
envelopes by Neri et al.\ (1998), which contains maps of 46 envelopes
in the CO (1--0) and (2--1) lines, made using the IRAM
telescopes. With respect to the structure of the envelopes, the
authors conclude that within the sensitivity of their observations,
AGB and post-AGB envelopes are for the most part spherical, with
evidence for an inner shell and/or significant asymmetry in 30\% of
the sample. In fact, most of these cases are in the former category,
so the truly asymmetrical envelopes are not very common. This is in
marked contrast to the widespread asymmetries in the ionized gas of
PNe, discussed throughout this volume.

\begin{figure} 
\plottwo{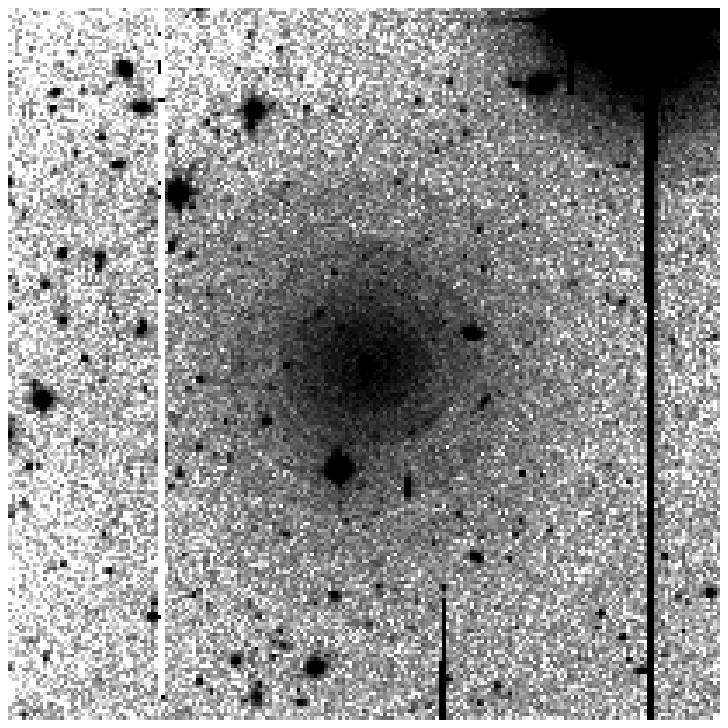}{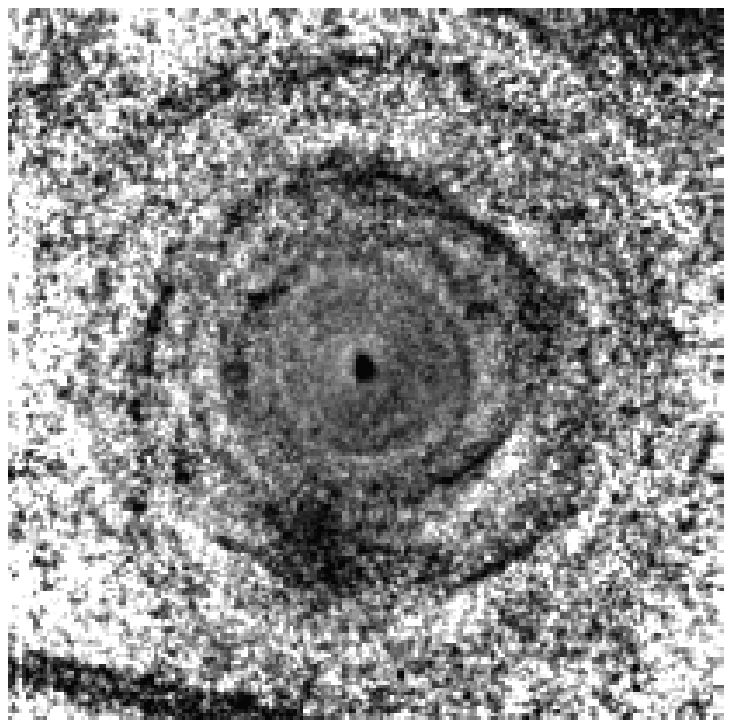}
\caption{Multiple shells in the envelope of IRC+10216. 
\emph{Left}: Deep $V$-band
image (field 223\arcsec). \emph{Right}: Detail of center 
with an averaged envelope profile and background objects subtracted 
(field 131\arcsec). Data from Mauron \& Huggins (1999). }
\end{figure}

One interesting question that arises in this context is: just how
regular and symmetric is the mass loss of a typical AGB star?  A good
example with which one can address this question is the archetype of
carbon-rich AGB stars, IRC+10216.  The molecular envelope of IRC+10216
has been intensively observed for more than a decade using millimeter
interferometry, in order to study the chemistry.  Recent results are
reviewed by Lucas \& Gu\'elin (1999).  A key point relevant to the
structure of the envelope, is that some molecular species are observed
to form more than one layer around the star, and the peaks of several
species coincide, which suggests the presence of an underlying
physical shell structure. About three distinct shells can be
identified within $\sim 25\arcsec$ of the star (see Fig.~4 of Lucas \&
Gu\'elin 1999).

This multiple shell picture of the envelope of IRC+10216 has recently
been extended at optical wavelengths (Mauron \& Huggins 1999). Deep
images obtained with the CFHT reveal the envelope of IRC+10216 in
dust-reflected, Galactic light. The left hand panel in Fig.~1 shows a
$V$-band image, with a field of 223\arcsec, and the the right hand
panel shows a close-up of the center, enhanced to show details.

The figures show that the envelope is not smooth, but consists of a
series of nested, limb-brightened shells. The inner shells show a
rough correspondence to those in the molecular gas.  The shells are
not complete in azimuth, and are separated by somewhat irregular
intervals that correspond to time scales of $\sim 200$--800~yr. This
structure appears to be a basic feature of the mass-loss process but
is not yet understood: the time scale is much longer than the stellar
pulsation period, but is much shorter than the interval between the
thermal pulses.

The presence of these shells in the archetype AGB envelope is
especially significant in the context of recent HST observations of
multiple arcs (with similar spacings) in a half-dozen proto-PNe and
young PNe (see Terzian, this volume). The number of cases found
already suggests that they may be common, and may well be the norm for
certain classes of objects. These arcs can almost certainly be
identified with the shells seen in IRC+10216 -- at a more advanced
stage of evolution -- so their origin can be traced back to the
formation of the molecular/dust shells in the precursor AGB envelopes.

\begin{figure} 
\plottwo{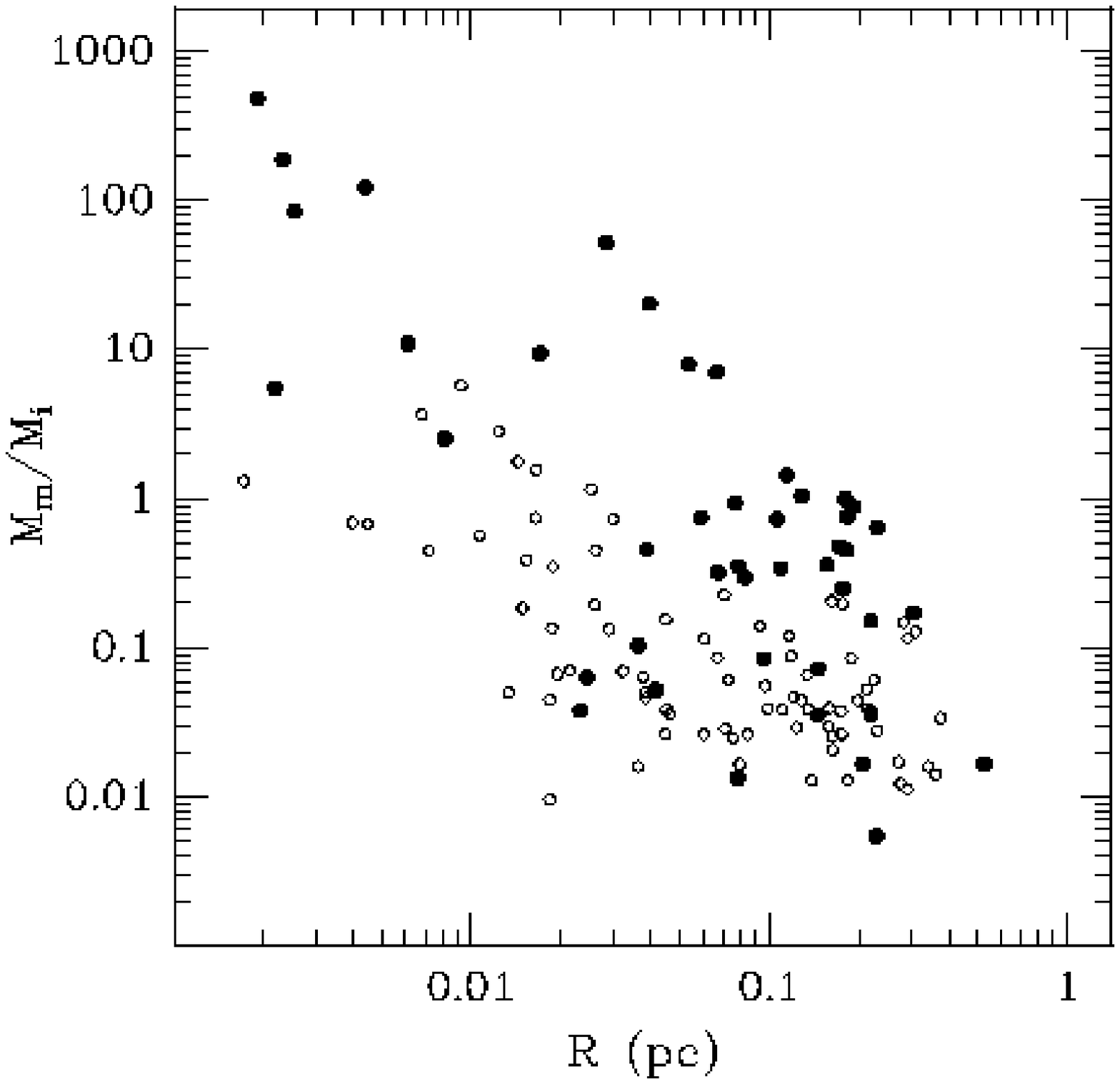}{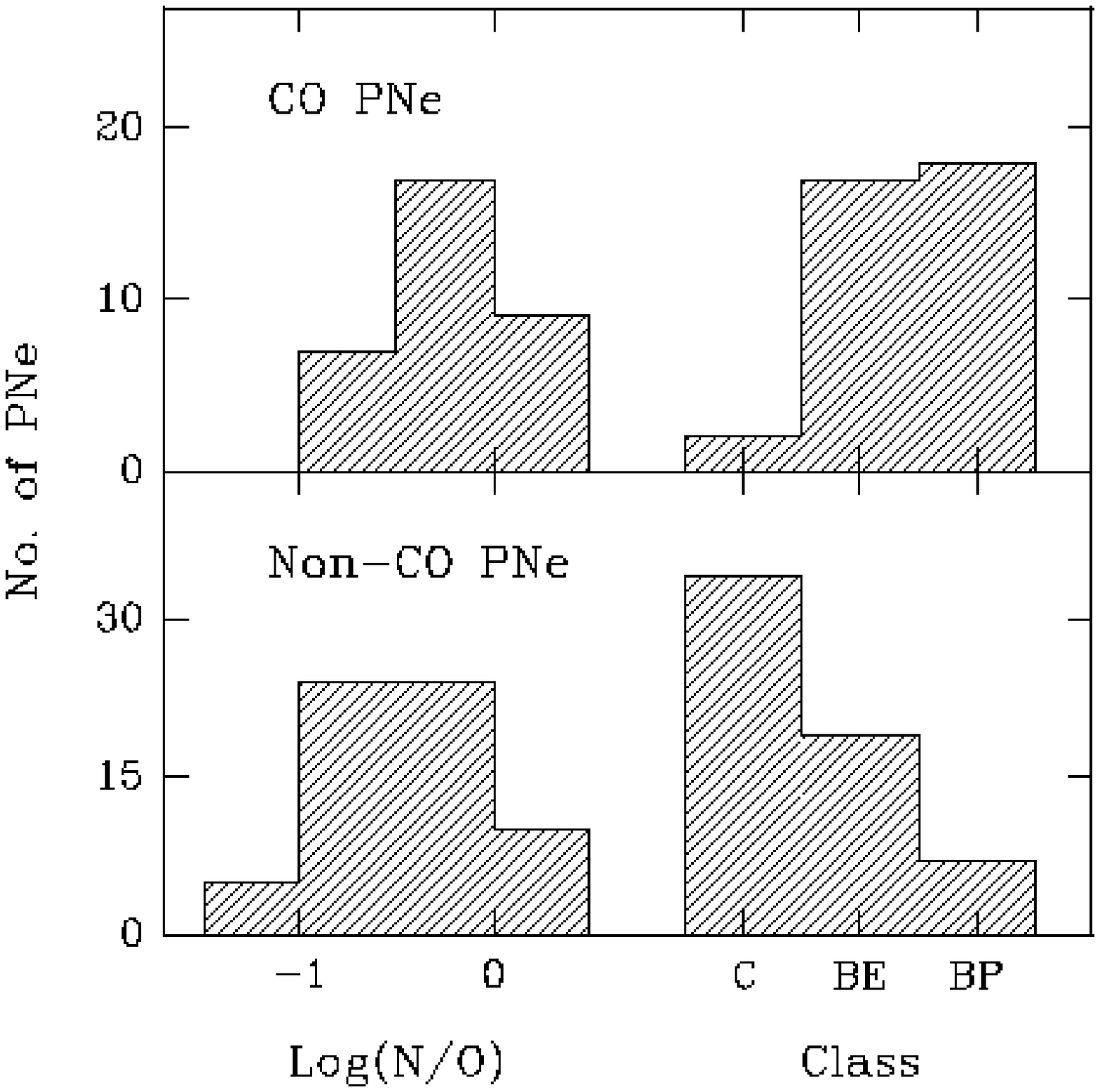}
\caption{Molecular gas in PNe. \emph{Left}: Mass ratio of molecular
to ionized gas vs.\ PN size; filled circles are detections, open
circles upper limits.  \emph{Right}: Correlation of CO detections with
the N/O abundance ratio and PN shape: centric (C),
elliptical (BE), bipolar (BP). Data from Huggins et al.\ (1996).}

\end{figure}

\section{Molecular Gas in PNe}
In spite of major changes in the mass loss and the onset of
photo-ionization through the post-AGB phase, a significant component
of molecular gas is found in many \emph{bone fide} PNe, even in highly
evolved cases. The most widely detected molecular signatures are
the 1.3 and 2.6~mm lines of CO and the 2~$\mu$m vib-rotational lines
of H$_2$.  Recent survey work has been reported by
Huggins et al.\ (1996) and Kastner et al.\ (1996) for CO and H$_2$,
respectively, with more than 40 PNe detected in each
species. Additional examples are continually being added to the lists
(e.g., Josselin et al.\ 1999; Hora \& Latter 1999).  Other molecular
species are also detected in the neutral gas, even in some evolved
cases (Bachiller et al.\ 1997), and various aspects of the chemistry
have recently been discussed by Natta \& Hollenbach (1998), and Howe
\& Williams (1998).

The molecular gas is found predominantly in PNe at low Galactic
latitude, and there is a strong correlation with morphology (e.g., see
Fig.~2, right panel), which suggests that we detect the nebulae with
higher mass progenitors.  There is no doubt about the location of the
gas. Except in the youngest PNe where the envelopes may still
completely enshroud the nebula, the molecular gas is found around the
waist of the ionized gas, in shapes variously described as rings,
cylinders, or toroids.

One example which illustrates the relation of the molecular gas to the
nebula is M2-9. This young ($\tau_{\rm exp} \sim 1,500$~yr), bipolar
PN is well known from imaging with HST. It is seen nearly sideways-on,
and is dominated by elongated, twin outflows or jets.  The H$_2$
emission in M2-9 lies along the edges of the cylindrical structure
formed by the jets (Hora \& Latter 1994). The CO emission, which
traces the densest gas, is found only in a slowly expanding
($V_{exp}$ = 7 km\,s$^{-1}$) torus which fits tightly around the waist
of the nebula (Zweigle et al.\ 1997).

\begin{figure} 
\plotone{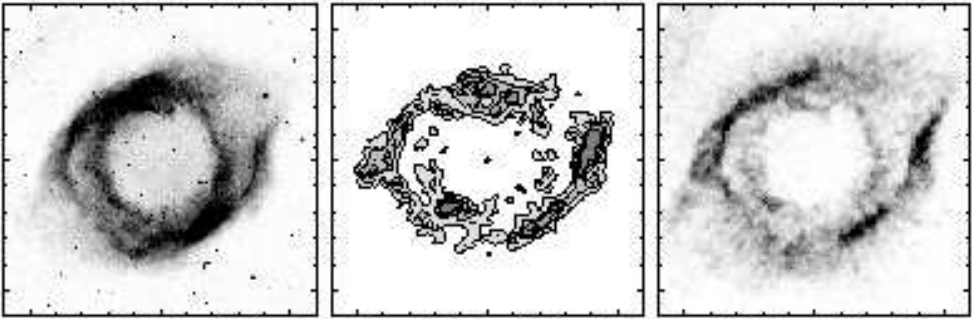}
\caption{Structure of the molecular envelope of the Helix nebula. 
\emph{Left}: $R$-band image. \emph{Center}: Integrated intensity map
in the 1.3~mm CO (2--1) line. \emph{Right}: ISO (LW2)
image, dominated by the $v=0-0$ S(5) line of H$_2$. Each field is 1200\arcsec.
Data from Young et al.\ (1999) and Cox et al.\ (1998).   }
\end{figure}

A second example is the Helix nebula (NGC 7293) which is a much older
system ($\tau_{\rm exp} \sim 10,000$~yr), seen nearly end-on. The
2~$\micron$ H$_2$ emission has been imaged by Kastner et al.\ (1996)
and the whole nebula has recently been mapped in the CO(2--1) line by
Young et al.\ (1999).  The left panel of Fig.~3 shows an $R$-band
image of the Helix (mainly H$\alpha$ and [N\,{\sc ii}]), and the next
panel shows the CO map. These two views are quite similar: it is
evident that the ionized nebula abuts the envelope and has formed
through photo-ionization of the neutral gas.  Thus, in both the Helix
and M2-9, and in many other PNe as well, the molecular gas is a main
structural feature of the nebula and an important key to its
morphology.

The mass of molecular gas in the PN envelopes is not easy to
determine, even though the major constituent (H$_2$) is directly
observed.  This is because the 2~$\micron$ emission arises from high
lying levels ($> 6000$~K) and typically samples only a small fraction
of the gas. (For recent work on whether shocks or ultraviolet
radiation excite the lines, see Hora \& Latter 1999). More useful mass
estimates come from the low lying lines of CO, which are thermalized
under a wide range of conditions. The CO fluxes can be used to
estimate the total number of CO molecules, and with reasonable
assumptions on the CO abundance lead to an estimate of the mass.
Masses estimated in this way range up to a few $M_{\sun}$. 

To illustrate the evolution of the envelopes, the mass ratio of
molecular to ionized gas in a large number of PNe is shown in Fig.~2
(left panel) vs.\ PN size -- which is a rough measure of the age.  The
upper envelope of this plot defines a striking evolutionary trajectory
for PNe with a significant molecular component (statistically the
bipolar PNe): the young PNe are dominated by the envelope, which
remains a significant component of the circumstellar gas until they
reach a size of $\sim 0.1$~pc. In other PNe, the molecular gas is
photo-dissociated more rapidly.

In addition to the molecular gas, there is, not surprisingly, ample
evidence for neutral atomic gas in PNe, e.g., from observations of the
21~cm line and infrared fine structure lines. This gas is in interface
regions between the molecular and ionized gas, and in envelopes that
are essentially completely atomic.  The masses in these components are
substantial (e.g., Taylor et al.\ 1990; Dinerstein et al.\ 1995; Young
et al.\ 1997), but they have not yet been studied in large numbers of
PNe or at high angular resolution, so they do not yet provide a
systematic or detailed picture.

\section{Large and Small Scale Structure in the Molecular Gas}
The detailed structure of the molecular gas in PNe is of considerable
interest since it contains information on the physical processes that
produce the nebulae. For space considerations, we focus here on one
example, the Helix nebula, which is among the nearest PNe and can be
examined with high spatial resolution.

One aspect of the structure in the gas is the high degree of
fragmentation. The cometary globules located within the inner, ionized
gas are well known from optical imaging with HST, and have been
discussed by O'Dell \& Burkert (1997).  Millimeter CO observations
have shown that the globules have dense cores of molecular gas
(Huggins et al.\ 1992); in fact without this structure it would be
hard to account for their presence in large numbers, because they
would be rapidly photo-ionized and disperse on short times scales,
compared to the age of the nebula.  Recent work by Meaburn et al.\
(1998) shows that the globules have kinematic similarities to the
inner CO ring seen in Fig.~3, and it seems likely that they share a
common origin with the more extended molecular envelope.

This envelope itself is also highly fragmented.  The clumpy structure
in the CO map in Fig.~3 (resolution 30\arcsec) is seen to consist of
many smaller clumps at higher resolution (see Huggins 1999).  The
masses of the clumps cannot be more than an order of magnitude more
than the cometary globules in the ionized gas, and they share
remarkably narrow line widths of $\sim 1$~km\,s$^{-1}$. These
fragments appear to be close cousins of the cometary globules, and
probably represent a slightly earlier stage in the development of
these structures.

This picture of a fragmented torus is confirmed by observations with
ISO (Cox et al.\ 1998).  Spectroscopy at 5--17~$\mu$m with ISOCAM
reveals a strong, pure ($v=0-0$) rotational spectrum of H$_2$ in the
Helix, from the S(2) to the S(7) lines.  As one moves from the ionized
cavity to the limb, the spectrum changes from nearly featureless, to
being dominated by the H$_2$ emission.  The line intensities indicate
an excitation temperature of $\sim 900$~K, thus the H$_2$ is warmer
than the CO, and probably forms a skin on the cooler gas.  The S(5)
line emission dominates the ISO LW2 filter so it has also been
possible to image the large scale distribution of H$_2$ over the whole
nebula.  The image, at 6\arcsec\ resolution, is shown in Fig.~3, right
panel. It shows finer details than the CO map, including flocculent
radial rays and clumps of globules around the inner periphery, and
generally underscores the completely fragmented picture of the
molecular gas.

A second aspect of the distribution of the molecular gas is its global
structure. Fig.~3 shows that it is not simply a regular torus, and the
apparent ``double ring'' seen in optical images has long been a target
for speculation. The large scale geometry has recently been discussed
by Young et al.\ (1999) using their CO observations.  The CO map in
Fig.~3 actually consists of 3425 spectra which record the velocity of
the gas along the line of sight as well as the intensity. Since the
system is expanding, these can be used to study the 3-dimensional
structure.

When examined in this way, the data indicate that the inner ring in
Fig 3, is a true ring, tilted $\sim 37\deg$ to the line of sight.  The
outer arcs to the east and west peel away from the ring (from the
north and south, respectively), with a remarkable degree of point
symmetry in their geometry (see Fig.~7 in Young et al. 1999).  For
each point in the structure, there is a similar one on the opposite
side of the central star.  There seems little doubt that this
structure, which dominates the nebula, was formed at an early stage by
the interaction of the envelope with collimated, bipolar outflows or
jets.

\begin{figure}
\plotone{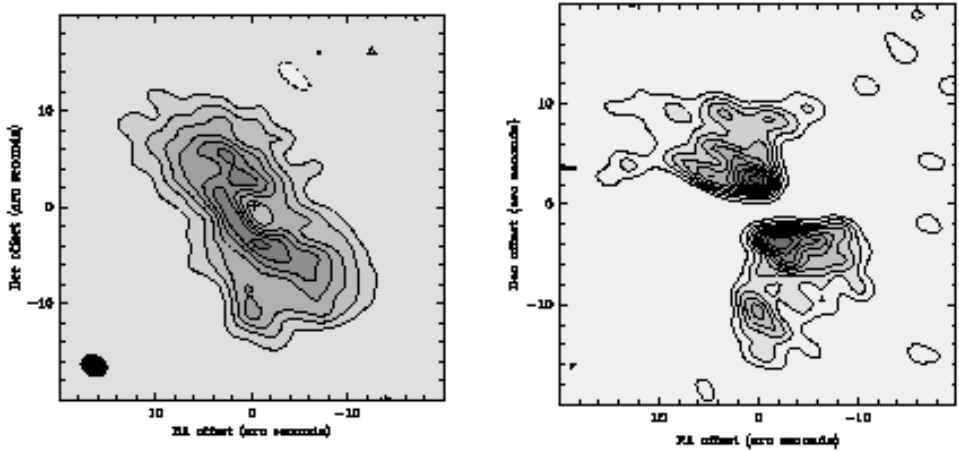} 
\caption{Molecular disk of the young PN KjPn~8.  \emph{Left}:
Integrated intensity map in the CO (1--0) line. \emph{Right}: Channel
map of the CO (1--0) emission at the LSR systemic velocity of
$-$35~km\,s$^{-1}$. Data from Forveille et al.\ (1998).}
\end{figure}

\section{Shaping the Envelopes}

The actual shaping of envelopes by collimated outflows or jets from
the central star system is well documented in a number of proto-PNe
and young PNe by high resolution observations of the molecular gas.
These observations are especially interesting in light of the growing
number of PNe that are seen to exhibit point symmetries (especially
young PNe observed with HST, see Sahai this volume). These symmetries
indicate that collimated outflows -- and their interactions -- are
common in young PNe, and can provide a general shaping mechanism.

The presence of a molecular (or atomic) envelope is an interesting
part of this scenario: it not only affects the dynamics of the
interaction, but the high densities and low temperatures of the
neutral gas preserves the results of the interactions for much longer
than the ionized gas. Thus at later stages, this structure can
dominate the appearance of the PNe, as is the case in the Helix
nebula described above.

Examples of outflow-envelope interactions in the proto-PN phase are
described elsewhere in this volume (e.g., by Lucas, and Alcolea), and
we focus here on young PNe. The most spectacular outflows are those in
KjPn~8 . These consist of pairs of bipolar jets, the most
recent of which have expansion velocities $\sim 200$~km\,s$^{-1}$ and
extend over $\sim 4\arcmin$, even though the ionized nebula is only
$\sim 4\arcsec$ in size (see L\'opez, this volume, for details). Using
the IRAM interferometer, Forveille et al.\ (1998) have recently mapped
the molecular emission from the KjPn~8 system (Fig.~4, left
panel). The molecular gas forms a disk that surrounds the ionized
nebula (which corresponds to the hole in the center of the figure),
and exceeds it in mass by a factor of $\sim 60$, so it dominates the
circumstellar environment.

The disk axis and the jets are aligned in KjPn~8, and their expansion
time scales are similar (a few thousand yr). A likely scenario is that
common (or related) mechanisms ejected the molecular gas and formed
the jets close to the central star. The jets and the gas clearly
interact. In the right panel of Fig.~4, the CO channel map at the
systemic velocity shows a wind swept disk, with gas extending along the
edges of the jet cavities; other channels show that this gas has the
highest velocities, and is entrained in the flows. In this case,
a primary result of the interaction is the shaping of the disk or
torus.

A second example of outflow-envelope interactions occurs in
He3-1475. In this young PN, high resolution optical imaging with HST
(Borkowski et al.\ 1997) shows high velocity, bipolar outflows with
large opening angles that are apparently focussed into narrow jets by
interactions with their surroundings. Recent CO observations with the
IRAM interferometer (Forveille, private communication) reveal a torus
of molecular gas around the outflows. The torus extends a few arc
seconds along the flows, roughly corresponding the region in which the
flows are collimated; much of the molecular gas is also at high
velocities and is entrained.  The interesting perspective here is that
the outflow-envelope interactions are complex: the molecular torus is
probably the focusing agent of the jets, while at the same time the
flows are acting to shape and disrupt the envelope. Further studies
of these types of shaping interactions are currently in progress.

\section{Concluding Remarks} 
The examples described here illustrate and underscore the important
role of molecular gas in the structure of PNe.
\begin{itemize}
\item In the equatorial tori that dominate the evolution
of bipolar and related PNe for much of their lives. 
\item As the dense cores of globules and other microstructures,
\item
As the  origin of multiple arcs.  
\end{itemize}
High resolution observations of the molecular gas are also beginning
to reveal details of the shaping of PNe by envelope interactions with
collimated outflows or jets at an early phase.

\

\acknowledgments It is a pleasure to acknowledge collaborative
programs with R.~Bachiller, P.~Cox, T.~Forveille, and N.~Mauron, which
form part of this paper.  This work was supported in part by NSF grant
AST-9617941.

\end{document}